\documentclass[prl,final,aps,twocolumn]{revtex4}

\usepackage{graphicx}

\begin{document}

\title{The intermediate problem for binary black hole
inspiral and the periodic standing wave approximation
}

\author{Benjamin Bromley}
\affiliation{Department of Physics,
University of Utah, Salt Lake City, Utah 84112}  
\author{Robert Owen}
\affiliation{Theoretical Astrophysics, California
Institute of Technology, Pasadena, CA 91125}
\author{Richard H.~Price} 
\affiliation{Department of Physics \& Astronomy and Center for 
Gravitational Wave Astronomy, University of Texas at Brownsville,
Brownsville, TX 78520}

\begin{abstract}
{\bf Abstract}
In calculations of the inspiral of binary black holes an intermediate 
approximation is needed that can bridge the post-Newtonian methods
of the early inspiral and the numerical relativity computations of
the final plunge. We describe here the periodic standing 
wave approximation: A numerical solution is found to the problem
of a periodic rotating binary with helically symmetric standing 
wave fields, and from this solution an approximation is
extracted for the physically relevant problem of inspiral with outgoing waves.
The approximation underlying this approach has been recently confirmed
with innovative numerical methods applied to nonlinear model problems.

\end{abstract}

\maketitle

The computational study of the inspiral of binary black holes is
important for the understanding of gravitational wave signals, and is
of inherent interest as a question in general relativity that can be
answered only with computation. It has therefore become the focus of
supercomputer codes that evolve Einstein's field equations forward in
time from initial conditions chosen to represent a starting
configuration of the inspiralling objects. The evolution codes,
however, typically become unstable on a timescale (set by the size of
the hole) short compared to a full orbit. Reliable calculations of the
final plunge are now feasible, the merger and  ringdown of the final black
hole fate of the system are handled well with perturbation
theory\cite{laz1}, and the early inspiral is well approximated with
post-Newtonian computations\cite{PN}. What cannot be handled well is
the intermediate phase of the inspiral, that late epoch during which
nonlinear effects are too strong for the post-Newtonian approximation,
and too many orbits remain for numerical relativity to be stable.

It has long been recognized that the basis of an approximation scheme
should be the slow rate of inspiral, the small ratio of the orbital
time to the radiation damping time\cite{det,det2}.  Through an
adiabatic treatment of the slow inspiral, such an approximation could
give answers about the radiation and rate of inspiral in the
intermediate epoch.  And when the rate of inspiral became too rapid,
the intermediate approximation could hand the problem off to numerical
evolution codes to do the final orbit and plunge, and could supply the
ideal initial data to those evolution codes. The need and the concept
for an intermediate approximation have been clear, but such an
approximation has not been easy to implement. Along with several
colleagues\cite{WKP,WBLandP,rightapprox,paperI} we have based an
approximation of slow inspiral on a numerical computation of no
inspiral. That is, we seek a numerical solution of Einstein's
equations for binary objects that are in circular periodic motion, and
whose ``helically symmetric'' fields rotate rigidly with the source
objects.

The universality of gravitation suggests that the unchanging motion of
such a system is not compatible with outgoing radiation, and this
intuitive suggestion is confirmed by the mathematics of the theory.
We, therefore, seek a helically symmetric solution for the sources
coupled to standing waves. In a linear theory standing waves would be
a superposition of half-ingoing and half-outgoing solutions.  In
linear theory, one could (though without motivation) solve the
standing wave problem. From the fact that solution is half the
superposition of the ingoing and outgoing solutions, and from the
relationship of the ingoing and outgoing solutions, one could extract
the outgoing solution.  The crux of our periodic standing wave 
method is that even for highly nonlinear binary inspiral fields there
is an ``effective linearity.'' The standing wave solution, to good
accuracy, is half the sum of the outgoing plus ingoing solutions
despite the nonlinearities. In general relativity, therefore, we
should be able to solve the standing wave numerical problem and extract an
approximation to the outgoing solution.

 It is important to understand why effective linearity can be correct
for inspiral. In the strong-field regions very close to the sources, 
the solution is  very insensitive to the distant radiative
boundary conditions (ingoing, outgoing, standing wave). In this
near-source region a superposition of half the ingoing and half the
outgoing solution gives a good approximation solution, because 
it amounts to averaging two samples of the same thing. In the wave zone where
the outgoing and the ingoing solutions are very different, the fields
are weak enough that nonlinear effects are negligible, and once again
we can superpose. The separation of the strong-field region from the
boundary-influenced region should be a clear separation unless the
sources are rotating very close to $c$, in which case the wave zone
will start just outside the sources. It is, however, not expected
that ultrarelativistic source motion can occur during the slow
inspiral epoch of motion.

We have recently been able to confirm effective linearity.  (Technical
details are given in \cite{eigenspec}.)  This confirmation has been
achieved with a  model problem, since the validity of effective linearity
can only be carried out in a model problem. In general relativity,
there will be no ``true outgoing'' solution available for confirmation
until numerical evolution codes are fully developed. In addition, the
numerical features of the helically symmetric standing wave
calculation pose new challenges very different from those of evolution
codes, and are best resolved in the simplest context possible.

Our model problem is a nonlinear scalar field coupled to point-like
sources in Minkowski space, and satisfying the field equation
\begin{equation}\label{fieldtheory} 
\Psi_{;\alpha;\beta}g^{\alpha\beta}
+\lambda F=\nabla^2\Psi-\frac{1}{c^2
}\partial_t^2\Psi+ F={\rm Source}\,.
\end{equation}
Here the source is taken to be two points of unit scalar charge in
orbit around each other at angular frequency $\Omega$, and at radius
$a$.  The velocity parameter for the system $\beta=a\Omega/c$ is taken
to be of order unity, representing the strong-field tight binary for
which post-Newtonian approximations are inadequate.  Our methods,
however, require that $\beta$ not be too close to unity. More
explicitly, our method works best if the radiation is quadrupole
dominated, and our calculations focus on values of $\beta $ from 0.3
to 0.5.  We expect that the approximation of slow inspiral will break
down before this assumption breaks down, and numerical relativity
evolutions will take over the job of tracking the last part of an
orbit and the subsequent plunge and merger.

The term $F$ contains the nonlinearity in our model theory, and we
have found the following form, with parameters $\lambda$ and $\Psi_0 $,
to be very useful:
\begin{equation}\label{modelF} 
F=\frac{\lambda}{a^2}
\frac{\Psi^5}{\Psi_0^4+\Psi^4}\,.
\end{equation}
A crucial feature of $F $ is that like the nonlinearities of general
relativity, it is very large near the sources, and becomes negligible
far from the sources.  The $\lambda$ multiplier allows us to vary the
strength of the nonlinear term, and the $\Psi_0$ parameter allows us
to vary the profile of the nonlinearity in the strong field region.

Our problem is defined by Eqs.~(\ref{fieldtheory}) and (\ref{modelF}),
and the source motion at angular frequency $\Omega$ in the equatorial
plane.  As described in spherical coordinates, helical symmetry can be
imposed on the solution $\Psi(t,r,\theta,\phi)$ by restricting to
solutions of the form $\Psi(r,\theta,\varphi) $, where $\varphi $ is
the comoving azimuthal coordinate $\phi-\Omega t $.  By restricting
the solution in this way, we have eliminated the possibility of
``evolution.''  For such helically symmetric solutions a change in
time by $\Delta t$ is the same as a change the azimuthal angle
$\Delta\phi=-\Omega\Delta t$. With this suppression of evolution, we
have eliminated the sorts of instabilities that develop in evolution
codes. But we have introduced new difficulties. 

These new difficulties can be seen immediately in the form of the
helically restricted nonlinear scalar equation

\begin{displaymath}
{\cal L}\Psi\equiv 
\frac{1}{r^2}\frac{\partial}{\partial r}
\left(r^2\frac{\partial\Psi}{\partial r}\right)
+\frac{1}{r^2\sin\theta}\frac{\partial}{\partial\theta}
\left(\sin\theta\frac{\partial\Psi}{\partial\theta}\right)
\end{displaymath}
\begin{equation}\label{ourL} 
+\left[\frac{1}{r^2\sin^2\theta}
-\frac{\Omega^2}{c^2}
\right]\,\frac{\partial^2\Psi}{\partial\varphi^2}
={\rm Source}-F(\Psi)\equiv\sigma(\Psi)\,.
\end{equation}
The principal part of this quasilinear equation is ``mixed,'' elliptic
inside a cylinder at $r\sin\theta=c/\Omega $, and hyperbolic outside
that cylinder. The problem is to be solved with radiative conditions
(ingoing, outgoing, or standing wave as described below) on a
spherical surface at large distances.  Well posed problems in physics
typically supply cauchy data on open surfaces to hyperbolic equations,
and Dirichlet or Neumann data on closed surfaces to elliptic
equations.  Our model leads to a boundary value problem with
``radiation'' conditions on a closed surface surrounding a mixed
problem. Though unusual, our problem is intuitively well posed, and
passes a computational test: we have found no fundamental difficulty
in solving models of this type numerically. Furthermore, a careful
analysis\cite{torre} of a closely related problem proves that
solutions exist and are stable.

``Standing wave'' solutions -- half ingoing and half outgoing --are at
the heart of our method, but there is not an obvious definition
of standing wave solutions in a nonlinear theory. Our procedure is to
find the outgoing ${\cal L}^{-1}_{out}$ and 
ingoing ${\cal L}^{-1}_{in}$ Green functions for Eq.~(\ref{ourL}). 
In principle, we can then iterate to find a solution of 
Eq.~(\ref{ourL}). The iteration 
\begin{equation}\label{outit} 
\Psi^{(n+1)}_{out}={\cal L}^{-1}_{out}\big[\sigma(\Psi^{(n)}_{std}
)\big]\,,
\end{equation}
if it converges, gives $\Psi_{out}$,
our nonlinear outgoing solution (and similarly for $\Psi_{in}
$), while  
the convergent result of 
\begin{equation}\label{stdit} 
\Psi^{(n+1)}_{std}
={\textstyle\frac{1}{2}} \left({\cal L}^{-1}_{out}+{\cal
L}^{-1}_{in} \right) \big[\sigma(\Psi^{(n)}_{std} )\big]\,
\end{equation}
is what we mean by our nonlinear standing wave solution, 
$\Psi_{std}$. The standing wave solution
$\Psi_{std}$
is fundamentally different from 
$(\Psi_{out}+\Psi_{in})/2$, but if effective linearity
is correct, the two  are very nearly equal.
(Note: In practice, for strong nonlinearities, the direct
iteration described above must be replaced by Newton-Raphson
iteration.)

\begin{figure}[ht] 
\includegraphics[width=.26\textwidth]{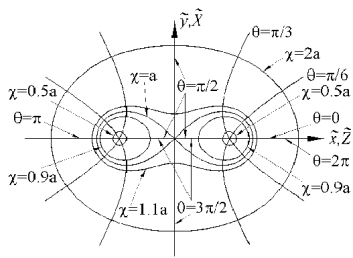}
\caption{
TCBC coordinates in the equatorial plane.
\label{fig:2dros}}
\end{figure}

We have previously\cite{paperI} solved the model problem of
Eq.~(\ref{ourL}) with more-or-less straightforward finite differencing
and direct matrix inversion. (The mixed nature of the partial
differential equations prevents the use of such efficient techniques
as explicit relaxaton.) This approach was successful (iterations
converged) for models with a limited range of source velocities and
nonlinearities.  More recently we have developed an innovative
numerical method that gives remarkably good results, with very little
computational cost, and that might be useful in problems other than
ours. Our new method is based on three elements.  First, we use
``adapted coordinates,'' comoving coordinates that conform to the geometry of
the problem. Near the source points our coordinate surfaces approach
those of source-centered spherical harmonics; far from the sources the
coordinates become spherical polar coordinates centered on the
midpoint of the orbit. Our specific choice of adapted coordinates is
two-centered bipolar coordinates $\chi,\Theta,\Phi$ (pictured in the
equatorial plane of the orbit in Fig.~\ref{fig:2dros}), which
asymptotically approach spherical coordinates $r,\theta,\phi$.  As
discussed in \cite{eigenspec}, this choice is not the most
computationally efficient possibility, but it has the advantage of
relative simplicity.

The second element of our method is to expand in multipoles (spherical
harmonics) of the adapted coordinates. If the separation between the
sources is much larger than the size of the source, then the only
feature of the source that is important is the ``local'' monopole
(monopole in source-centered coordinates).  In the wave zone, on the
other hand,  only the quadrupole of the field matters for the
radiation (unless the source motion is  highly relativistic). This
suggests that a multipole expansion in the adapted coordinates need
retain only the monopole and quadrupole. This turns out to be true for
source speeds $\leq0.3\,c$. For somewhat larger speeds, good accuracy
requires that the hexadecapole be kept in addition. This severe
filtering of the multipoles reduces the computational burden of
solving the problem, but more important, it eliminates the numerical
noise at short angular scales that we found for straightforward finite
difference computations.

An ``eigenspectral'' treatment of multipoles is the third element of
our method that is innovative. 
A straightforward approach to
dealing with multipoles would be to use $Y_{\ell m}(\Theta_i ,\Phi_j
)$ on an angular grid of the $\Theta$ and $\Phi $ coordinates, that
is, to use the multipoles of the continuum mathematics evaluated on
the discrete numerical grid. We have found that this approach does not
work at all well for our purposes. Because the monopole is so much
greater than the quadrupole in the wave zone, projecting out the
quadrupole component with the continuum multipole gives a large error.
We have therefore used the multipoles that seem ideally suited to
decomposition on the angular grid with $n_\Theta\times n_\Phi$ grid
vertices.  We view the values of our solution as vectors
$\Psi(\Theta_i, \Phi_j)$ in a space of dimension $n_\Theta\times
n_\Phi$. At large $\chi$ the angular Laplacian
\begin{displaymath}
L\equiv \frac{1}{\sin\Theta}\frac{\partial}{\partial\Theta}\left(
\sin\Theta\,\frac{\partial}{\partial\Theta}
\right)+\frac{1}{\sin^2\Theta}\,\frac{\partial^2
}{\partial\Phi^2}
\end{displaymath}
can be implemented as a finite difference operator, on the
$n_\Theta\times n_\Phi$ space, that is self-adjoint with respect to
the finite difference equivalent of integration over solid
angle\cite{nakamura}.  The eigenvectors of this self-adjoint operator
are approximately
$Y_{\ell m}(\Theta_i ,\Phi_j)$, but the eigenvectors are exactly
orthogonal so that the projection of the radiative multipoles is not
affected by the large monopole.

\begin{figure}[ht] 
\includegraphics[width=.22\textwidth]{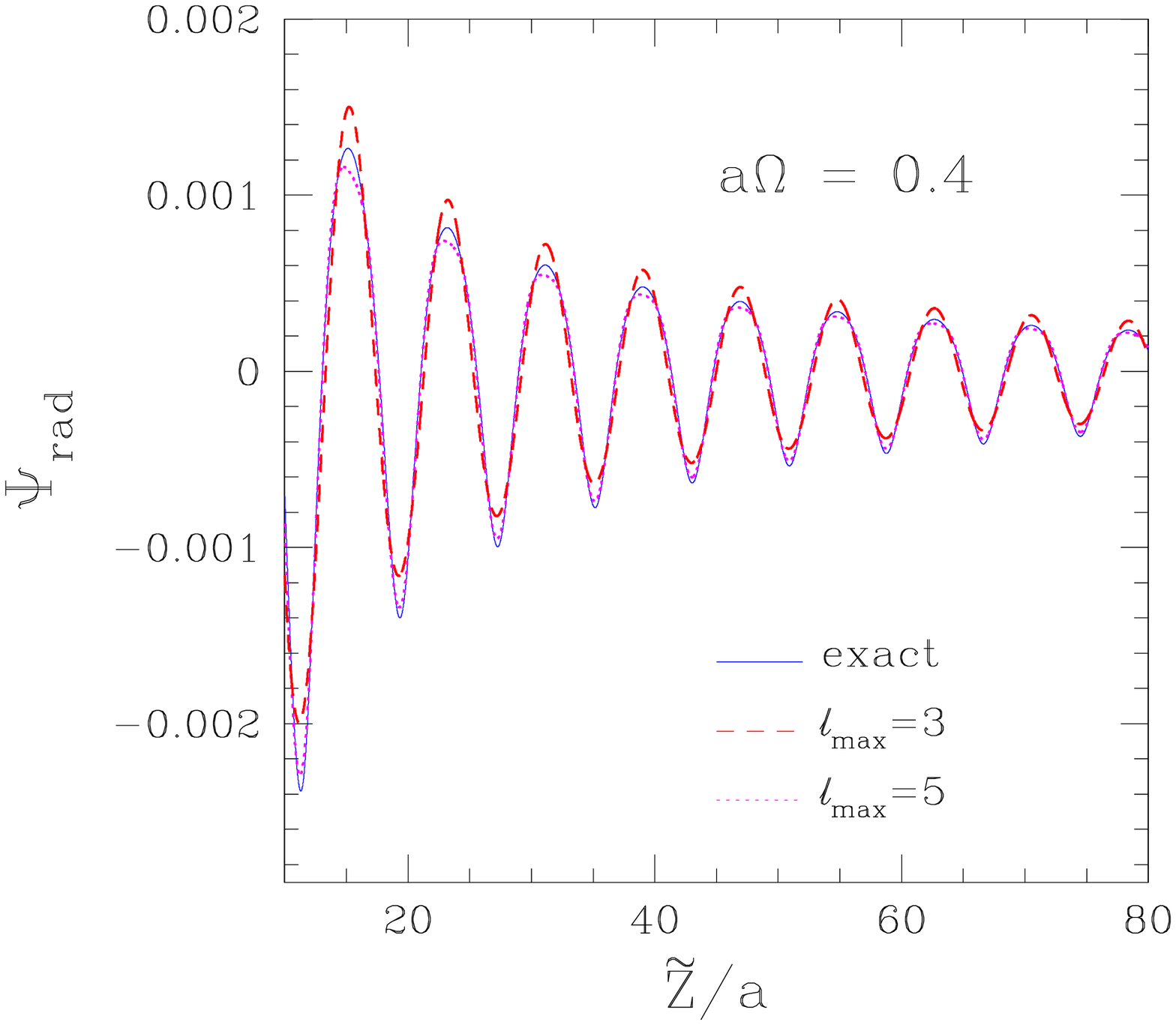}
\includegraphics[width=.22\textwidth]{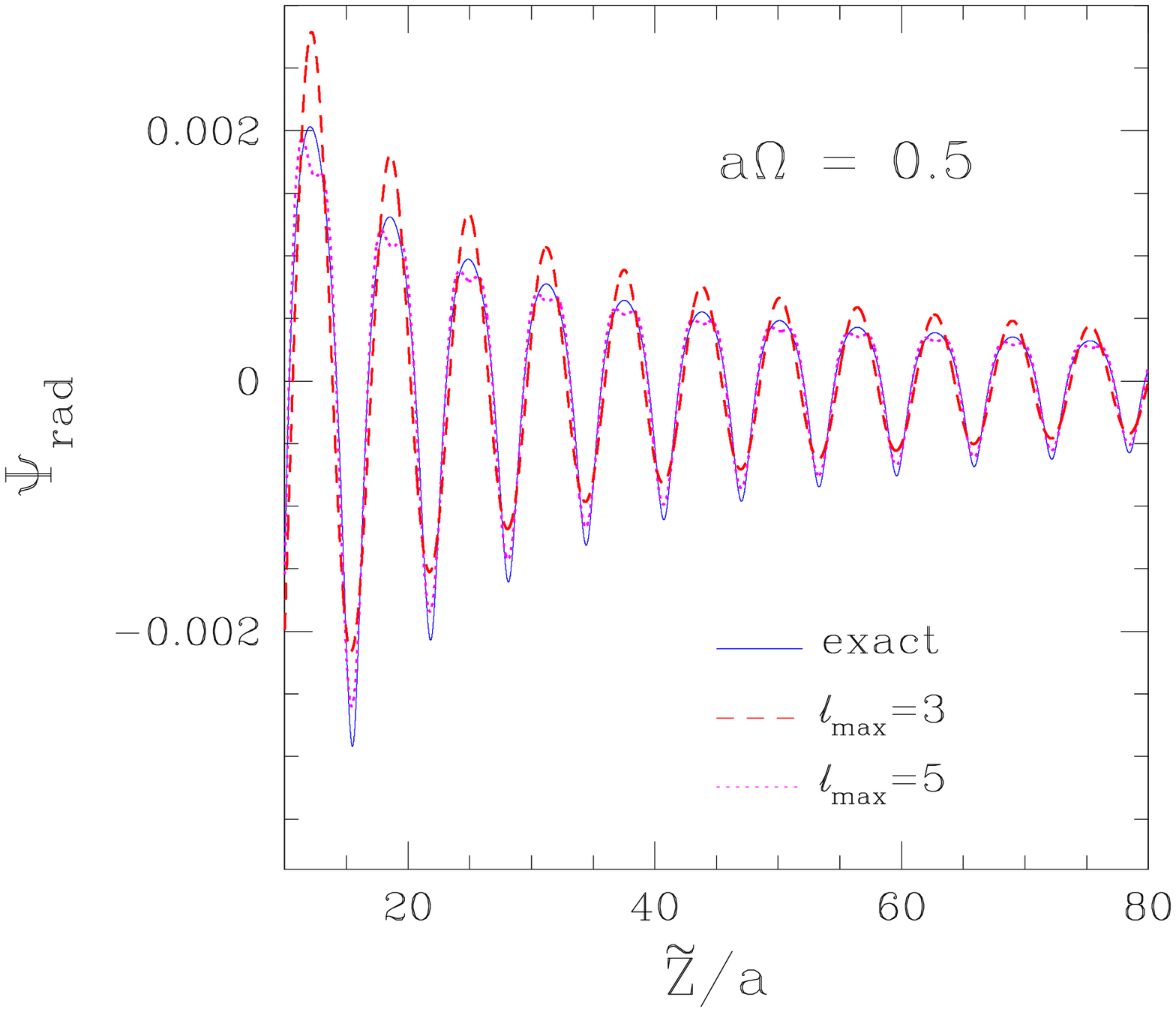} \caption{A comparison of
exact (series) linear outgoing solutions with eigenspectral 
solutions. The radiation field ($\Psi
$ minus its monopole) is plotted against the comoving coordinate $\widetilde{Z}
$, the distance from the center along a line through the source points. 
Eigenvectors were found on a $40\times80
$ grid for one quadrant, and 16001 radial grid points were used, with an inner
boundary condition at 
$0.02\,a$, and a Sommerfeld outer condition at  $80\,a$.
\label{fig:compare}}
\end{figure}

Questions about the validity of our eigenspectral method apply just as
much to the linear version as to the nonlinear version of our model
problem, and for the linear problem exact (infinite series) solutions
exist for comparison.  Figure~\ref{fig:compare} shows such
comparisons for
models with source velocity $a\Omega=0.4c$ and $a\Omega=0.5c$
and shows multipole filtering that allows either the monopole plus
quadrupole ($\ell_{max}=3$), or with the hexadecapole included
($\ell_{max}=5$). As the figure indicates, and as should be expected,
the number of multipoles that need be included to achieve a given
accuracy increases with increasing source speed. The minimal
$\ell_{max}=3$ multipole set may be adequate for approximate results
at $a\Omega=0.4c$ and gives excellent results for $a\Omega=0.3c$ (not
shown). There is, of course, no need to limit the number of multipoles
retained to such a small set, but as the number of multipoles
approaches the maximum number of eigenvectors that can be found for
the angular grid, the smoothing effect of multipole filtering is lost
and numerical noise becomes significant. A large number of multipoles
would be necessary only for source speeds $a\Omega$ close to $c$. As
already mentioned, it is unlikely that such motion will satisfy the
slow inspiral condition for which our approximation is designed.

\begin{figure}[ht] 
\includegraphics[width=.23\textwidth]{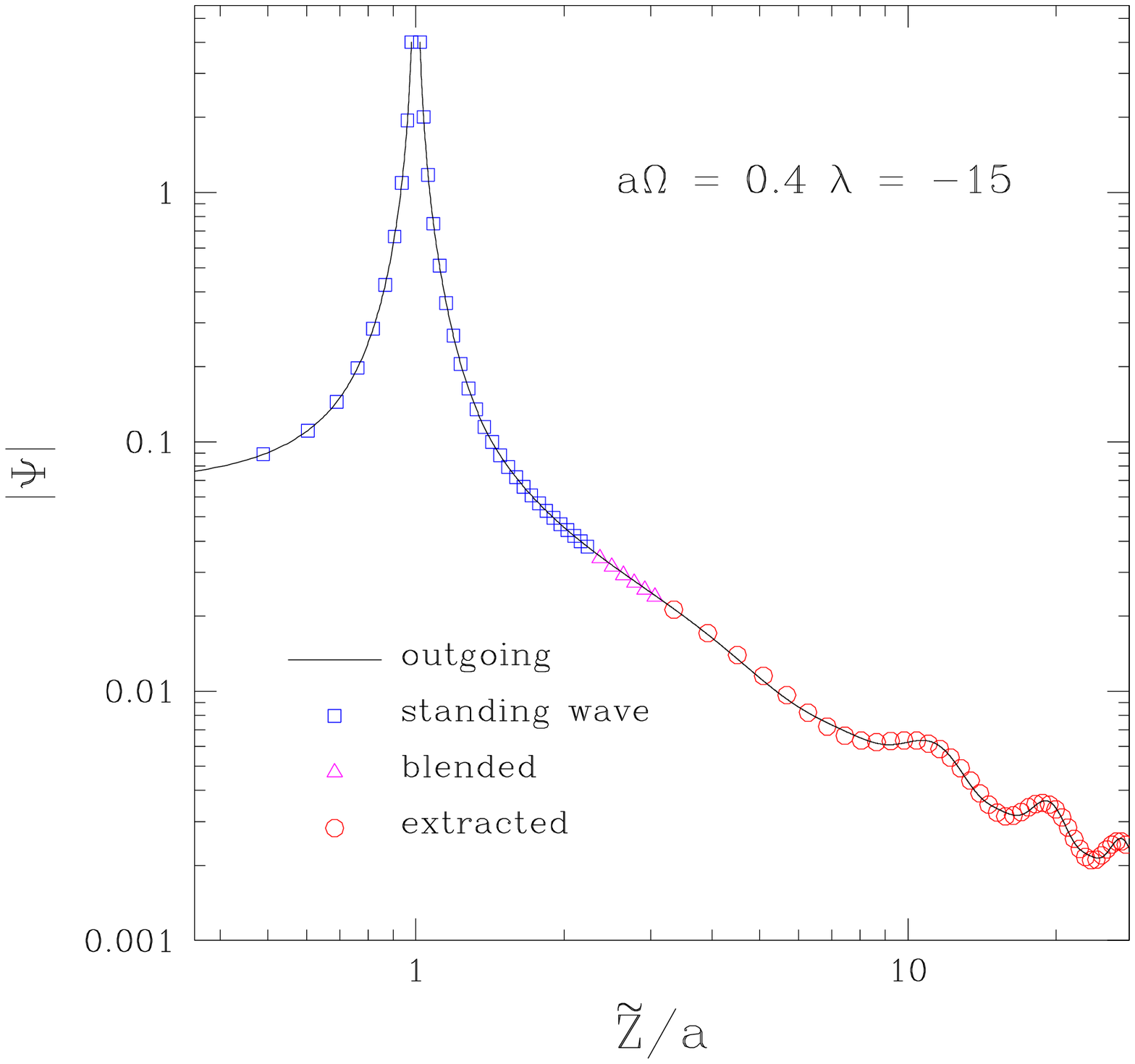}
\includegraphics[width=.23\textwidth]{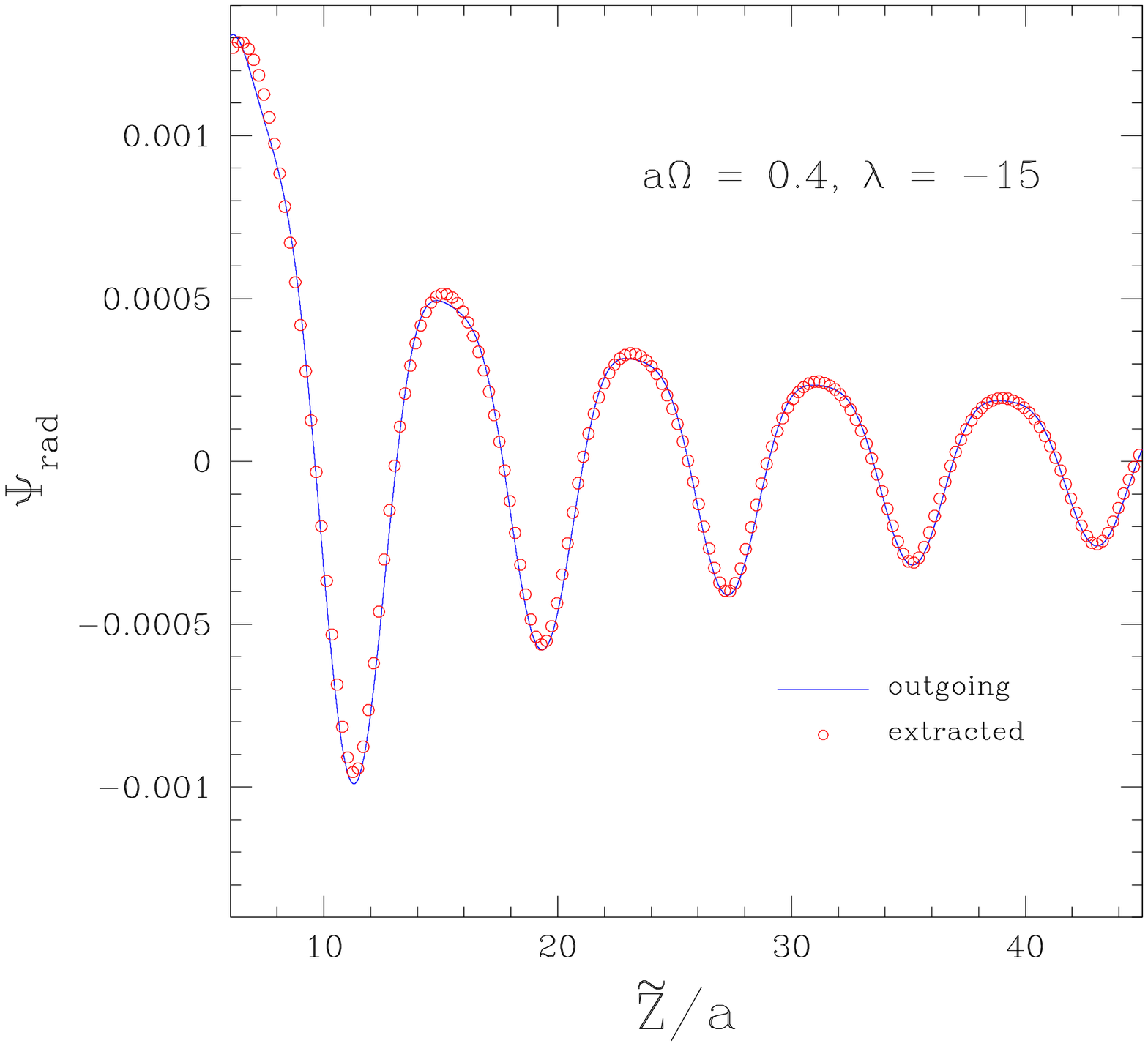} \caption{
The computed nonlinear outgoing solution compared with an approximate
outgoing solution extracted from the computed nonlinear standing wave
solutions. Grid parameters are the same as for Fig.~\ref{fig:compare},
and $\ell_{\rm max}
=5
$ was used.
\label{fig:extract}}
\end{figure}

Other confirmations of the validity of the method, especially for
nonlinear models, have been carried out, and are reported in
Ref.~\cite{eigenspec}. We focus here on the most important question
that can be answered with these models and numerical methods: Does
effective linearity work? Can we extract a good approximation to the
outgoing nonlinear problem from the sort of standing wave computation we
will be limited to when dealing with Einstein's theory?
Figure~\ref{fig:extract} gives strong evidence that we can.
For the chosen  parameters $\lambda=-15 $ and
$\Psi_0=0.15$, nonlinearities are significant, strong enough to reduce field
strength by around two-thirds. The outgoing and standing wave
solutions were each  computed by the Newton-Raphson version of the iteration
in Eqs.~(\ref{outit}),(\ref{stdit}).  

An outgoing approximation is extracted from the standing wave solution
by the following steps: (i)~In the outer region the solution is
matched to a general linear standing wave solution of half-ingoing and
half-outgoing waves; an outgoing wave is extracted as if the problem
were linear. (ii)~In the strong-field region very close to the source,
the standing wave solution itself is taken to be the outgoing
approximation.  (iii)~The two solutions are blended over a narrow
intermediate range of radii.

In Fig.~\ref{fig:extract}, the computed outgoing nonlinear solution is
shown as a solid curve. The data-type points representing the outgoing
wave extracted from the standing wave solution show how good the
approximation is. We have run models with much stronger nonlinearity
and have found equally good, or better, agreement of the true outgoing
solution and the extracted approximation. The validity of effective
linearity should, in fact, become questionable not for stronger
nonlinearity, but only for physically implausible high source velocity.

In addition to confirming effective linearity, computation with the
 model has also allowed some early insights about sensitivity to
 source details. By varying the multipole content of the inner boundary data
 we  explored the impact of source structure on the radiation
 field. The result (detailed in Ref.~\cite{eigenspec}) is in perfect
 accord with physical intuition; the radiation is insensitive to
 source structure unless the source size becomes comparable to the
 separation of the sources (i.e.\,, unless the moments ascribable to
 the structure of the individual sources are comparable to the
 quadrupole moment due to the separation of the mass points). The
 equivalent question for Einstein's theory is more difficult, but we
 should be able to give clear quantitative answers.

The next steps in our program start with linearized gravity in the harmonic
gauge. We have already done this with a finite difference code; work
on applying the eigenspectral method is underway, and no significant
problems are anticipated. The infrastructure of the linearized gravity
will provide much of what is needed for post-Minkowskian computations,
since the structure of the linear operators (the analogs of the ${\cal
L} $s in Eqs.~(\ref{ourL})--(\ref{stdit})) will be the same as for
linearized gravity. The final step to the full Einstein theory will,
in the same way, be based on the numerical infrastructure developed
for post-Minkowsian models.

We gratefully acknowledge the support of NSF grant PHY0244605 and NASA
grant ATP03-0001-0027, to UTB and of NSF grant PHY-0099568 and NASA
grant NAG5-12834 to Caltech.
We have greatly benefited from suggestions with John Friedman, Christopher
Beetle, and Lior Burko.


\begin{thebibliography}{10}
\bibitem{laz1}
J. Baker {\it et~al.}, Phys. Rev. D {\bf 65},  124012  (2002).

\bibitem{PN} L.~Blanchet, 
Living Rev. Relativity 5, (2002), 3. [Online article]: cited Dec.~25,
2000 http://www.livingreviews.org/lrr-2002-3.



\bibitem{det}
J.~K. Blackburn and S. Detweiler, Phys. Rev. D {\bf 46},  2318  (1992).


\bibitem{det2}
S. Detweiler, Phys. Rev. D {\bf 50},  4929  (1994).


\bibitem{WKP}
J.~T. Whelan, W. Krivan, and R.~H. Price, Class. Quant. Grav. {\bf 17},  4895
  (2000).

\bibitem{WBLandP}
J.~T. Whelan, C. Beetle, W. Landry, and R.~H. Price, Class. Quant. Grav. {\bf
  19},  1285  (2002).

\bibitem{rightapprox}
R.~H. Price, Class. Quant. Grav. {\bf 21},  S281  (2004).


\bibitem{paperI}
Z. Andrade {\it et~al.}, Phys. Rev. D {\bf 70},  064001  (2004).


\bibitem{friedmanetal} 
M. Shibata, K. Uryu, J. L. Friedman, Phys. Rev. D {\bf 70},  044044 (2004).
Erratum-ibid. D {\bf70} 129901
(2004).

\bibitem{torre} 
C.~G.~Torre, J.~Math.~Phys., {\bf44} 
6223-6232 (2003).

\bibitem{eigenspec} 
B. Bromley, R. Owen and R.~H.~Price, submitted 
to Phys.~Rev.~D. Preprint gr-qc/0502034.

\bibitem{nakamura}
After completing our scalar work with the eigenspectral method, we
discovered that essentially the same technique was used by
T. {Nakamura}, Progress of Theoretical Physics {\bf 72}, 746 (1984).

\end{thebibliography}
\end{document}